\documentclass[fleqn,10pt]{wlscirep}
\usepackage[utf8]{inputenc}
\usepackage{todonotes}
\usepackage[T1]{fontenc}

\newcommand\vartotalpackages{91,437}
\newcommand\varparsedpackages{74,829}
\newcommand\varclonedpackages{68,239}
\newcommand\varuniqueurls{51,657}
\newcommand\varuniquecloned{46,895}
\newcommand\varclonedgh{44,893}
\newcommand\varclonedgl{1,498}

\newcommand\varcommitsnobots{5,656,407}
\newcommand\varbotsgh{219}
\newcommand\varghusers{58,329}
\newcommand\varglusers{450}
\newcommand\varnoghglusers{14,266}
\newcommand\vartotalemails{89,399}
\newcommand\varmatchingghglusers{87}
\newcommand\varclonedothergit{504}

\newcommand\varghtreposght{13,623}
\newcommand\varghtreposrust{16,069}
\newcommand\varghtcommitsght{1,299,414}
\newcommand\varghtcommitsrust{1,942,997}
\newcommand\varghtusersght{14,563}
\newcommand\varghtusersrust{21,989}

\title{Evolving collaboration, dependencies, and use in the Rust Open Source Software ecosystem}

\author[1,$\dag$]{William Schueller}
\author[2,1,$\dag$]{Johannes Wachs}
\author[1]{Vito D.~P.~Servedio}
\author[1,3,4,*]{Stefan Thurner}
\author[5,6,1]{Vittorio Loreto}
\affil[1]{Complexity Science Hub Vienna, A-1080 Vienna, Austria}
\affil[2]{Vienna University of Economics and Business, A-1020 Vienna, Austria}
\affil[3]{Medical University Vienna, A-1090 Vienna, Austria}
\affil[4]{Santa Fe Institute, Santa Fe, USA}
\affil[5]{Sony Computer Science Laboratories, 75005 Paris, France}
\affil[6]{Physics Department, Sapienza University of Rome, 00185 Rome, Italy}

\affil[*]{Corresponding author: stefan.thurner@meduniwien.ac.at}
\affil[$\dag$]{These authors contributed equally to this work}

\begin{abstract}

Open-source software (OSS) is widely spread in industry, research, and government. OSS represents an effective development model because it harnesses the decentralized efforts of many developers in a way that scales. As OSS developers work independently on interdependent modules, they create a larger cohesive whole in the form of an ecosystem, leaving traces of their contributions and collaborations. Data harvested from these traces enable the study of large-scale decentralized collaborative work. We present curated data on the activity of tens of thousands of developers in the Rust ecosystem and the evolving dependencies between their libraries. The data covers eight years of developer contributions to Rust libraries and can be used to reconstruct the ecosystem's development history, such as growing developer collaboration networks or dependency networks. These are complemented by data on downloads and popularity, tracking dynamics of use, visibility, and success over time. Altogether the data give a comprehensive view of several dimensions of the ecosystem.
\end{abstract}
\begin{document}

\flushbottom
\maketitle

\thispagestyle{empty}

\section*{Background \& Summary}

Open Source Software (OSS) has recently been described as the ``infrastructure'' of the digital society~\cite{eghbal2020working}. OSS is an excellent example of open collaboration among many individuals that has a significant impact on the economy~\cite{lerner2002some, greenstein2014digital,nagle2018learning,nagle2019open}. Within specific OSS \textbf{ecosystems} - collections of software programs or libraries are in many cases, but not always delineated by the use of a particular programming language like Rust, Python, or PHP - developers contribute software that depends on software already in the ecosystem, often created by strangers. For instance, a library that generates data from probability distributions may use a random number generator from another library rather than writing a new one. The outsourcing of core functions leads to a \textit{rich structure of technical dependencies}, often represented as a network \cite{decan2019empirical}. These libraries are usually hosted on collaborative coding platforms like GitHub or GitLab.

The nature of OSS contributions is such that the \textit{traces of activity of individuals} are observable, i.e., what they contributed to which libraries and when. The cumulative efforts of thousands of developers can reveal a great deal about the nature of collaborative projects and work~\cite{zoller2020topology}. Information on the \textit{use}, \textit{visibility}, and \textit{popular success} of individual libraries can be tracked over time~\cite{sinatra2016quantifying}, along with the co-evolution of technical dependencies and social collaboration~\cite{cataldo2008socio}. Such data can give insight into the dynamics of massive and decentralized collaborations \cite{decan2019empirical} and how these digital ecosystems evolve.

Here, we present a comprehensive dataset on one such ecosystem built around the Rust programming language. Rust, a relatively young language, has recently seen a sharp increase in popularity. Besides its significant connections with Mozilla~\cite{jung2021safe}, it is, as of December 2021, the second approved language of the Linux kernel besides C. For several years now, it has been voted the ``most loved'' language in the Stack Overflow Developer Survey. We have collected and curated temporal data on the technical dependencies, developer contributions, and the use and success of individual libraries. Specifically, we can observe when a developer made an elemental contribution of code to a specific library, what other libraries that library depends on, and how widely used and popular the library is. We record over five million distinct contributions of over 72 thousand developers, contributing to over 74 thousand libraries over eight years. 

Our data processing pipeline, available as open-source software, combines data from Cargo (the Rust ecosystem library manager) and the code hosting platforms GitHub and GitLab. It considers and handles multiple issues common to the study of collaborative software development data \cite{kalliamvakou2016depth}: contributor disambiguation \cite{fry2020dataset,gote2021gambit}, bot detection \cite{golzadeh2021ground}, and the identification of nested projects and merged work. The result is a database tracking the evolution of a large, interconnected software ecosystem at a fine scale.

In contrast to other data sources on collaborative software development, our dataset contains more accurate and complete data for the Rust software ecosystem. Focusing on Rust allows us to integrate developer contributions with data on software dependencies and usage. In this way, our data is richer and more focused than what can be found in more extensive databases such as GHTorrent~\cite{gousios2012ghtorrent}, GHArchive, Software Heritage~\cite{pietri2019software}, or World of Code \cite{ma2019world}. Moreover, as we highlight in the Technical Validation section, we achieve a broader coverage by focusing on the Rust ecosystem: 15\% of the packages in our dataset are not in the GHTorrent database. Our dataset also requires significantly less storage space than the sources mentioned above and can be directly analysed by researchers with minimal computing infrastructure requirements. At the same time, Rust is a large ecosystem that has evolved in a decentralized manner with contributions from thousands of developers hosted on multiple platforms, differentiating it from data sourced from single projects like the Linux Kernel or Apache projects \cite{roberts2006understanding}. 

We plan to update the dataset annually provided that the primary upstream sources (crates.io, GitHub \& GitLab) remain stable. Researchers can also use our pipeline to reconstruct the dataset, a process that requires some data storage space (around 300 GB, though this volume will likely increase over time as the dataset is updated) and several days to query the data sources and process the results. Our code can also be adapted to collect data from other ecosystems, such as the Julia programming language's ecosystem. However, we note that not all ecosystems offer the same scope of data as Rust. For example: Rust is relatively young compared to Python, Ruby, or Javascript - as a result a much larger share of the Rust ecosystem's history is accessible on GitHub. The Rust ecosystem is also relatively small: estimates of the NPM (Node) ecosystem's size suggest its metadata alone are greater than 100 GB, while the repos themselves would take up multiple terabytes  as described here: \url{https://socket.dev/blog/inside-node-modules}.

We proceed as follows: first, we describe our data collection and wrangling process and the resulting database. We compare our data coverage against GHTorrent, a widely used database of OSS contributions, finding that our data is more complete. We then outline usage notes for researchers interested in topics such as online cooperation~\cite{zoller2020topology,szell2010measuring} and collaborative innovation~\cite{monechi2019efficient}, success~\cite{klug2016understanding}, and supply chain networks~\cite{ma2018constructing,zimmermann2019small,ohm2020backstabber} in software. Our data can easily be represented as, for example, dynamic networks of collaborating developers, time series of usage statistics, growing networks of interdependent libraries, or combinations thereof.

\begin{figure}[t]
    \centering
    \includegraphics[width=0.85\textwidth]{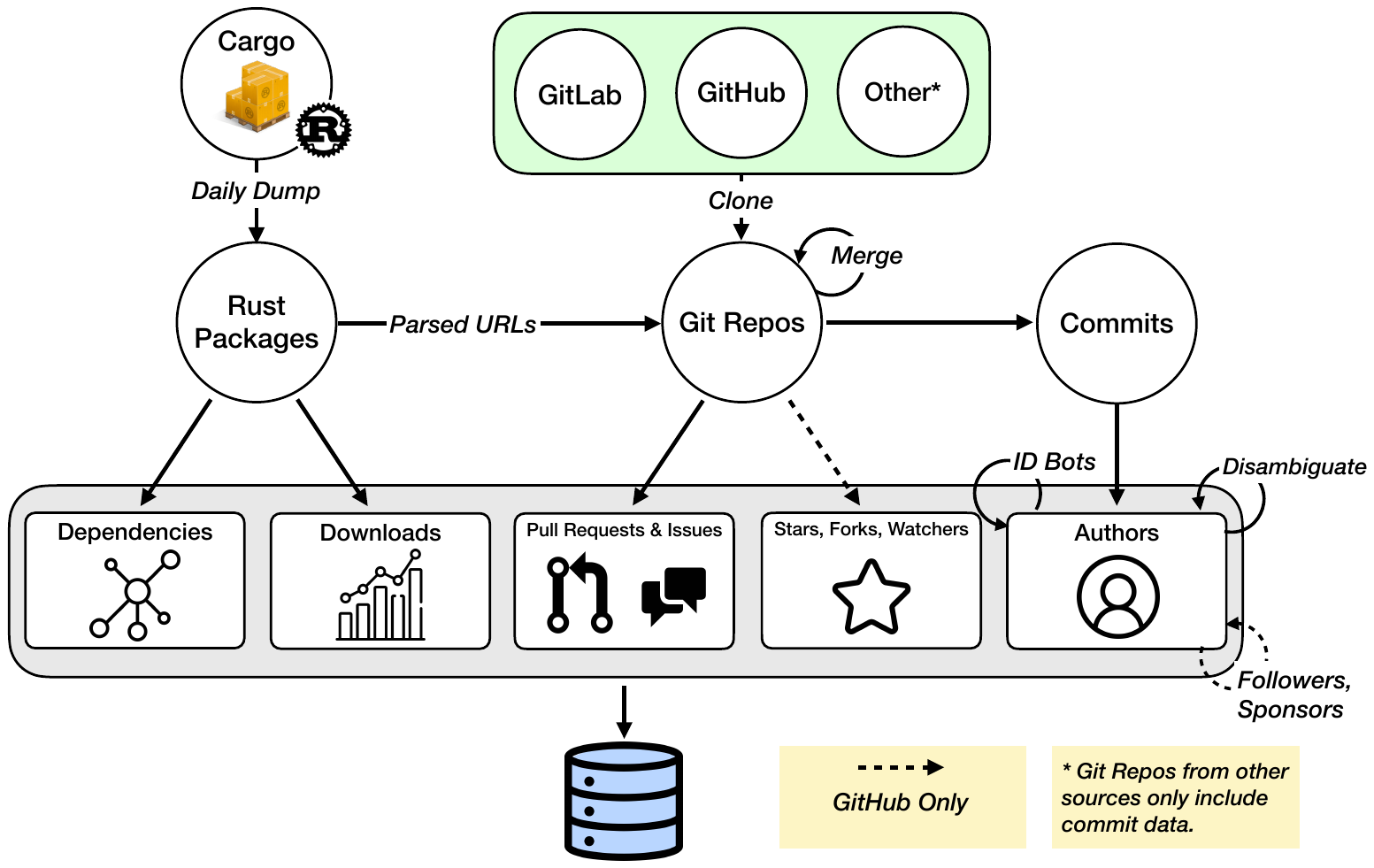}
    \caption{Data processing pipeline. We collect data from Cargo, the package registry of the Rust programming language, and complement it with data from the code hosting platforms GitHub and GitLab. The processed result integrates information on package dependencies, use (downloads and stars), and authors.}
    \label{fig:diagram}
\end{figure}

\section*{Methods}
We describe the data sources, and how we combine and curate data from various sources to create a comprehensive overview of the Rust ecosystem. We provide a visual overview of the established data processing pipeline in Figure~\ref{fig:diagram}.

\subsection*{Data Sources and Collection}

\subsubsection*{Cargo: Libraries and Dependencies}

Our first source of data is the Cargo package (which are called \textit{crates} in the Rust community) registry. Registries, often called package managers, play an important role in nearly all OSS ecosystems. They allow users to download and update different libraries while resolving dependencies and managing conflicts. Other examples of registries around different programming languages include PyPI for Python, CRAN for R, Rubygems for Ruby, and NPM for Node. We use Cargo as a source of technical dependencies and downloads for Rust. These are available as part of a daily dump from \url{crates.io}, accessible via: \url{https://crates.io/data-access}. 

The data can be directly imported in a PostgreSQL database, and contains package names and creation dates, their versions, a list of dependencies for each version with the semantic versioning (semver) syntax associated to them, and the daily downloads per version of each package. For a relevant discussion of the importance of semantic versioning in OSS ecosystems, see recent work by Decan and Mens~\cite{decan2019package}. Packages are also often associated to a repository URL and a documentation URL, but those are not always provided and depend on maintainer input.

\subsubsection*{Code Repositories: GitHub, GitLab and other git platforms}
To understand who contributes to which library, we turn to the social coding platforms on which these packages are hosted. In the case of Rust, nearly all packages in Cargo are hosted on either GitHub or GitLab. Specifically, \varparsedpackages packages had links to either platform. Of these links, \varuniqueurls\ were unique and \varuniquecloned\ of them could be cloned from either GitHub (\varclonedgh), GitLab (\varclonedgl) or other git platforms (\varclonedothergit). The inclusion of data from GitLab represents an important extension over the most widely used databases in OSS research, GHTorrent and GH Archive, which only use data from GitHub. Both of these platforms use Git version control, making projects hosted on either alternative comparable.

The elemental code contributions to OSS projects using the Git version control system are called \emph{commits} and are associated to email addresses belonging to the author and contributor (in practice these are usually the same). GitHub and GitLab both host Git projects (called repositories or \emph{repos}, for short), which we downloaded and used to extract information about activity and collaboration. The mapping between repos and the libraries hosted on Cargo is not one-to-one and requires additional processing, described below. The Git version control history of a project allows us to examine in detail the contribution histories of all developers working on a project. Indeed, previous work has shown how this highly granular data can be exploited to study collaboration and interactions among developers \cite{scholtes2016aristotle,gote2019git2net}. To do so, we ``clone'' (download) each repo locally. We also make use of the GitHub and GitLab GraphQL APIs to disambiguate contributors.

\subsubsection*{Measuring Use, Visibility, and Success: Downloads and Stars}
We quantify two dimensions of the use of libraries: the number of times they are downloaded and the number of times they received positive social feedback (stars) on GitHub. The two metrics highlight different aspects of use. Downloads, sourced directly from Cargo, present a more technical measurement of use. GitHub's stars are more suggestive of visibility. For example, a highly technical library that provides background functions may be downloaded many times but have relatively few stars. The social aspect of GitHub and platforms is known to play an important role in collaborative software engineering \cite{dabbish2012social,marlow2013impression}. GitHub stars and other forms of social feedback including followers and sponsors matter as information signals in the software development community for both libraries and contributors \cite{borges2018s,moldon2021gamification}, for example on the labour market \cite{papoutsoglou2019extracting}. Of course, library owners can and do seek to raise GitHub stars for their libraries by promoting them \cite{borges2018s}; stars are by no means an objective measure of library popularity. That said, we collected time-stamped GitHub stars and forks, a measure of code reuse, for repos directly from the GitHub GraphQL API. For GitHub users, we also collected the network of followers and sponsors each user has using the same API. GitHub implemented sponsorships for developers in 2019 \cite{shimada2022github}, enabling developers to crowdfund from users who appreciate their work directly on their GitHub pages.

\subsection*{Data Processing}
Several additional curation steps need to be taken to insure the data collected is useful for the purposes of research into digital collaboration. We describe three specific steps here.

\subsubsection*{Repositories, Packages, Forks}
One would easily assume individual packages of the Rust ecosystem tracked by Cargo correspond one-to-one with repositories on GitHub or GitLab, and this is generally the case. However, some repositories host several packages. For example modules or plugins that extend a core package are often hosted together in the same repo, but are distinguished in Cargo. This presents a data challenge: dependencies are recorded between packages, while contributions are recorded at the repo level.

For each package listed in the registry, one or several URLs are typically provided. They correspond to a link to the code and/or the documentation. Here, we take the URL corresponding to the code, and if empty, coalesce it with the documentation URL. We parse the individual URLs by recognizing the prefixes synonyms of \textit{github.com} and \textit{GitLab.com} and the pattern \verb|github.com/<owner>/<name>| corresponding to a repository. 

We also try to identify other git platforms by matching the pattern \verb|<platform_root_url>/<owner>/<name>|, and using \verb|git ls-remote| on a maximum of 5 repos per candidate to check if it is indeed a git platform.

Repositories are sometimes renamed, and both URLs can be present in distinct packages. The old URL typically redirects to the new one. To resolve this and be able to merge repositories under one entity, we use the \verb|<owner>/<name>| returned by the GraphQL API (of both GitLab and GitHub) when querying about the repository, for example when collecting information about forks.

After downloading the repositories (also called cloning), we analyse the commit data and retrieve commit hash, author and contributor emails and names (usually the same), and commit parents. We also compute the number of lines added and deleted for each commit. We analyse all available branches, and the unique hash ensures that we do not count commits twice per repository. However, commits can appear in several repositories, when one repository is a fork of another. We keep attribution of commits to each repository where they appear, but we also attribute each commit to a main repository, supposed to be its origin. For this, we retrieve the information about forks from the GitHub or GitLab GraphQL API, and take as origin the highest repository in the fork tree containing the commit. When this method is unsuccessful (e.g., undeclared forks, or forks between different Git platforms), we take the repository having the oldest package, using its creation date from the package manager.

\paragraph*{Collaborative Coding Events}
For repos hosted on GitHub and GitLab, we also collect data on collaboration events including issues and pull requests. Developers can open issues, highlighting bugs and problems with a code base. Maintainers can respond in comments and close issues, indicating whether they have been addressed. Pull requests are how non-core developers contribute code to a project - these can be commented upon and merged into a project, or rejected. These events, comments on them, their labels, status, and emoji reactions to them are all recorded in our data with time stamps. Similar data (comments and emoji reactions) for commits on GitHub are provided as well. Seeking to preserve developer privacy, we do not provide the texts associated to these events.

\subsubsection*{Dependencies}
When analysing the dependencies between packages sourced from Cargo, and aggregating the network to dependencies between repositories, we noted the presence of cycles. In this context a cycle represents a pattern like: Package A depends on Package B, Package B depends on Package C, and Package C depends on Package A. Though there were only few of such examples, we decided to prune dependencies to remove such cycles for two reasons. The first is that they represent a logical inconsistency in what a dependency means. Second, without cycles, the resulting dependency network is a directed acyclic graph (DAG). DAGs are themselves interesting data structures appearing in a variety of data science contexts~\cite{corominas2013origins}. Given the small number of packages involved in cycles, we manually inspected them. Dependency cycles in package space correspond mainly to unnecessary dependency links or even fake packages for the sake of testing dependency declarations. Repository space cycles are more complex to prune. We adopted the heuristic to remove the dependencies (in both repo and package space) in cycles of length 2 by pruning the dependency of the oldest node to the newest (by creation date or earliest date of the repo's corresponding packages), and naturally removing dependency cycles of length 1. The remaining cycles were inspected manually, and the cycles were broken by removing the dependency links where it made more sense, in most cases from a repository having one of the highest download counts to one having one of the lowest. One remaining repository, although corresponding to numerous downloads, has been pruned from all dependencies to it because of the high number of cycles of the dependency network involving it. We included these pruned dependencies in the dataset for the sake of completeness, but we flagged them for easy removal, or for letting the possibility to investigate other link removal policies. We guarantee absence of cycles for the state of the dependency network at the end of the dataset (2022-09-07) and at the end of the preceding year (2021-12-31) in both spaces, but not at any arbitrary timestamp.

\subsubsection*{Merging of Developer Identities}
There are many potential ways to disambiguate the identities of contributors~\cite{fry2020dataset,gote2021gambit}, each presenting tradeoffs. In general, Git commits are signed by an email address, not a platform-specific username. Developers often commit code from different computers or environments with different email addresses in their configuration files, and this can result in a significant disambiguation problem. Rather than attempting to infer which email addresses potentially refer to the same person, we query the GitHub API for the GitHub account linked to each commit \cite{montandon2019identifying}. While this ignores the potential that developers use multiple accounts, we argue that it makes a larger amount of highly justifiable merges among commit author identities than an email address based approach. Email address-based author identity disambiguation would scale better in larger systems at the cost of accuracy.

For each email address, we carry out this process for the most recent commit registered for that email, and if this fails to return an account, we try again with a randomly sampled commit among all those corresponding to this email. After doing the same for GitLab, we also merge matching GitHub and GitLab logins (finding \varmatchingghglusers such cases). An additional step is to parse the emails and spot those obviously belonging to GitHub, following the patterns \verb|<login>@users.noreply.github.com| or \verb|<randomint>+<login>@users.noreply.github.com|.

\subsubsection*{Bot Accounts}
Bots play an important role in modern software development \cite{wessel2018power,wessel2022bots}, but need to be handled with care in any study of software systems, as they can make orders of magnitude more contributions than any human developer. Pooling them and their activity with that of human developers would skew any analysis of social cooperation and collaboration \cite{kalliamvakou2016depth,golzadeh2021ground}. While bots have interesting effects on project evolution \cite{wessel2018power}, we chose to detect and mark bots and more generally invalid accounts in our dataset with a view to excluding them as we are primarily interested in the patterns of contributions of developers. To identify bots, we used a two-step filtering process. First, we extracted all bots on a curated list used as ground truth for bot detection in the software engineering community \cite{golzadeh2021ground}. We then filtered remaining GitHub accounts with the substrings ``*[bot]'', ``*-bot'',``*-bors'',``bors-*'',``dependabot-*'' in their usernames, and finally inspecting manually each individual account with pattern ``*bot*''. After filtering out bot accounts labelled this way, a manual inspection of the 100 most active remaining accounts was conducted, as well as of all accounts containing the substring ``bot''. The manual inspection took the following steps: checking the GitHub webpage of a user for a clear name, looking at the description, looking at the commit/PR comments. The 100 most active  accounts by number of commits across the full timespan of the dataset were considered, as well the 100 most active accounts in the year 2020.

For users that could not be associated to a GitHub account, their emails are filtered when the last part of their prefix (separated by ., - or +) is equal to ``bot'', ``ghbot'', ``bors'', ``travis'' or ``bot''.  A few remaining email strings without ``@'' were discarded, like ``localhost'', ``N/A'' or empty string. Manual inspection of the most active 100 emails without a GitHub account revealed a few more bots. The list of manually discarded bots is available in the file \verb+botlist.csv+. Our dataset includes the bots among the users, flagged with a Boolean ``is\_bot'' attribute to enable filtering.

\section*{Data Records}
We host our data on Figshare \cite{ws2022data} and the code used to collect and process the data from our sources on GitHub (\url{https://github.com/wschuell/repo_datasets}). Both platforms track the history of the data and code, allowing researchers to use any version they prefer as we continue to update and extend both. We share data in several formats, noting that in no format does the data exceed 6 GB when compressed.

\begin{table}[!t]
\begin{tabular}{ll} 
\setlength\tabcolsep{2.5pt}
Table Name & Description\\
\hline
\verb+commit_comment_reactions+ & Individual emoji reactions to commit comments \\
\verb+commit_comments+ &  Comments to commits \\
\verb+commit_parents+ & Parenthood relationships between commits. Typically one per commit, can be 0 or more. \\
\verb+commit_repos+ & Repos to which commits belong. At least one, but can be several (e.g. with forks). \\
\verb+commits+ & Listing metadata about specific commits \\
\verb+filtered_deps_package+ & Packages wich are filtered when appearing as a dependency to avoid cycles \\
\verb+filtered_deps_packageedges+ & Dependency edges between packages filtered to avoid cycles \\
\verb+filtered_deps_repo+ & Repositories wich are filtered when appearing as a dependency to avoid cycles \\
\verb+filtered_deps_repoedges+ & Edges directly discarded in the dependency graph \\
\verb+followers+ & Listing followers of GitHub accounts \\
\verb+forks+ & Listing forks declared on GitHub \\
\verb+identities+ & Listing each individual identity of the developers (email, GitHub account, Gitlab account) \\
\verb+identity_types+ & Listing of identitiy types (email, GitHub account, Gitlab account) \\
\verb+issue_comment_reactions+ & Individual emoji reactions to issue comments \\
\verb+issue_comments+ & Comments to issues \\
\verb+issue_labels+ & Individual labels of each issue \\
\verb+issue_reactions+ & Individual emoji reactions to each issue \\
\verb+issues+ & Listing of issues per repository \\
\verb+merged_identities+ & Keeping track of identities having been merged, for information purposes \\
\verb+merged_repositories+ & Repositories having been merged (after identifying renaming or typo in URL) \\
\verb+org_memberships+ & Membership of organization declared on GitHub for GitHub users. \\
\verb+organizations+ & Organizations or work groups as declared on GitHub \\
\verb+package_dependencies+ & Listing package dependencies (version to package with semver) \\
\verb+package_version_downloads+ & Listing daily downloads of package versions \\
\verb+package_versions+ & Listing versions of packages \\
\verb+packages+ & Listing packages \\
\verb+pullrequest_comment_reactions+ & Individual emoji reactions to each pull request \\
\verb+pullrequest_comments+ & Comments to each pull request \\
\verb+pullrequest_labels+ & Individual labels of each pull request \\
\verb+pullrequest_reactions+ & Individual emoji reactions to each pull request \\
\verb+pullrequests+ & Listing of pull requests per repository \\
\verb+repo_languages+ & Listing language composition of repositories (GitHub) \\
\verb+repositories+ & Listing repositories \\
\verb+sources+ & Listing data sources \\
\verb+sponsors_user+ & Listing sponsorships of developers \\
\verb+stars+ & Listing starring events of repositories \\
\verb+urls+ & Listing retrieved URLs, and their parsed equivalent \\
\verb+user_languages+ & Language composition of contributions of GitHub users (year before collection). \\
\verb+users+ & Listing users (who can have several identities: email, GitHub, Gitlab) \\
\verb+watchers+ & Listing watchers (developers being notified of changes) of each repository \\
\hline

\end{tabular} 
  \caption{Table of tables in the database. A full database schema is available on Figshare: \url{https://doi.org/10.6084/m9.figshare.c.5983534.v1} and on \url{https://github.com/wschuell/repo_datasets}.}
  \label{tab:table_of_tables}
\end{table} 

Before we describe the data, we discuss data pseudonymisation. To preserve developer privacy, we provide data that is scrubbed of information that can directly be used to identify individuals. We do so in the following way: we discard name attributes and hash (via MD5 with a random salt) email addresses and GitHub/GitLab logins. Researchers interested in studying social or demographic characteristics of developers, such as gender \cite{vasilescu2015gender,rossi2022gender}, geography \cite{rastogi2018relationship,braesemann2019global,wachs2022geography}, or both \cite{prana2021including}, could adapt our approach to data collection and analyse these attributes. However, they should consider potential ethical issues that arise when associating such information to users \cite{gousios2017mining}.

In Table~\ref{tab:table_of_tables}, we list the tables in the database along with a description of their content and purpose. For the sake of brevity, we refer the reader to the accompanying materials on Figshare \cite{ws2022data} for a schema of the database and a description of the variables included.

\section*{Technical Validation}

In this section, we report statistics on the completeness of our data. An advantage of defining the Rust ecosystem as all those packages hosted on Cargo, is that we can precisely measure how many packages we can successfully integrate into our database. In particular, we can report the share of packages that we can connect to repos on the social coding platforms GitHub and GitLab. As we will see below, we have a very high rate of linkage. Moreover, the packages that we could not integrate are typically those with very few downloads. This suggests that most projects on Cargo that do not appear on GitHub or GitLab are small personal projects or preliminary work.

\subsection*{Package Coverage among Repositories}
\begin{figure}[t]
    \centering
    \includegraphics[width=0.8\textwidth]{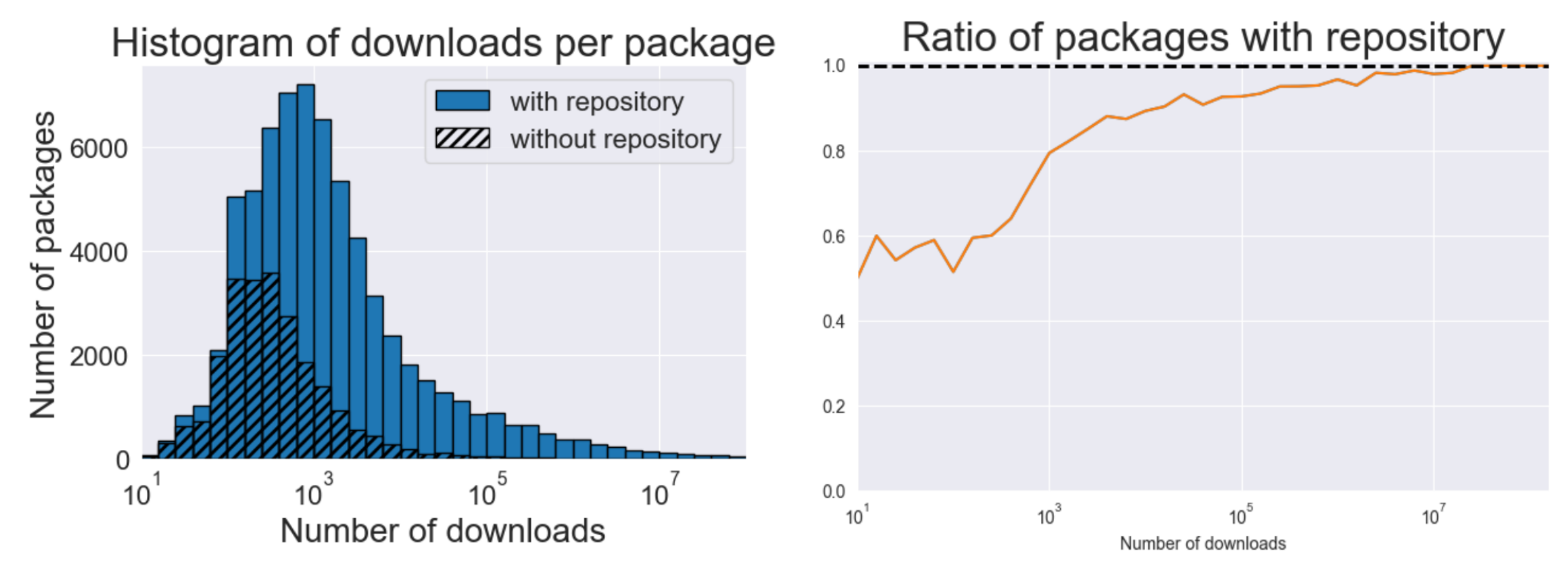}
    \caption{Data coverage: we check the number of Rust packages for which we could identify and download a corresponding Git repo (from GitHub or GitLab) in terms of their use, measured in downloads. We could link a large majority of packages to repos, and have a significantly higher success rate if we consider packages that have been downloaded more often.}
    \label{fig:completeness}
\end{figure}

Some Rust packages hosted on Cargo could not be linked to repos on GitHub or GitLab. They either are on a different platform, for example on Bitbucket, Google Cloud, Sourceforge, or on personal websites. Still others had a link to a GitHub or GitLab domain (i.e. a repo) but could not be cloned. This can happen if a link was incorrectly transcribed, if the repo was deleted, or if the package is listed only for name squatting or test purposes and does not correspond to any repository. Specifically, out of \vartotalpackages\ 
 packages, \varparsedpackages\ were linked to a repository on GitHub, GitLab or another git platform and \varclonedpackages\  of these were successfully cloned, although only \varuniqueurls\ packages pointed to distinct URLs (and \varuniquecloned\ were cloned). Hypothesizing that the most important packages are the most downloaded ones, we can see in Figure \ref{fig:completeness} that our coverage increases among the most important packages, measured by use (downloads).

Across the cloned repos, we gathered \varcommitsnobots\ commits, the elemental units of contribution in the git version control system. Excluding \varbotsgh\ bots identified among the GitHub accounts, these contributions were made by \varghusers\ GitHub users and \varglusers\ GitLab users. The raw data contains \vartotalemails\ identifying email addresses, highlighting the significant amount of disambiguation of author identities our pipeline implements. \varnoghglusers\ of them could not be associated to a GitHub or GitLab account.

\subsection*{Comparison with other data sources}
We first compare our data with data collected in the GHTorrent project \cite{gousios2012ghtorrent}. The GHTorrent project aims to collect all activity on GitHub for use in research. As we have already noted, most activity in the Rust ecosystem takes place on GitHub, with a small but significant share taking place on GitLab. Besides the inclusion of GitLab data, we observed that our data contains a significant amount of activity hosted on GitHub that is missing from the GHTorrent database (the SQL version), when comparing the data in our dataset tagged as happening before the last date of user creation in the GHTorrent database -- May 31st 2019, just before midnight -- and corresponding to the repositories that could be cloned. Specifically, we found only \varghtusersght\ unique users (vs.\ \varghtusersrust\ GitHub users in our database), \varghtreposght unique repos (vs. \varghtreposrust\ identified GitHub repos in our database), and \varghtcommitsght\ unique commits (vs. \varghtcommitsrust\ GitHub hosted commits in our database before the last date of GHTorrent). The GHTorrent project uses the GitHub REST API and collects data from the public event timeline using user-donated API keys. Outages on either the GHTorrent side or on the GitHub REST API, or rate limited API keys may explain missing data. While GHTorrent remains an excellent source of dataset for all of GitHub, these comparisons suggests that for a focused look at a single ecosystem, a customised pipeline can significantly increase data coverage. More detailed statistics concerning the comparison can be found on Figshare \cite{ws2022data} in the file \verb+ghtorrent_comparison.yml+.

We also note that our data collection pipeline is not the only way to collect similar data on OSS ecosystems. For example, the GrimoireLab toolchain is a collection of tools to gather and analyze data on software \cite{gonzalez2022software}. These tools provide users with sophisticated analyses of the health and activity levels of particular projects, and groups of projects. In particular its Perceval data retrieval module can do many of the same things as our collection pipeline. Our tool focuses rather on one thing: collecting a nearly complete dataset on the Rust ecosystem quickly (i.e. maximizing GraphQL API calls vs. REST API calls), with minimal requirements for users seeking to replicate or refresh the dataset. In this sense the two can be considered as potential complements, rather than substitutes: an analyst studying specific libraries in the Rust ecosystem can easily use the GrimoireLabs to obtain metrics on those libraries and additional data about them, for example from Jira or Twitter.

\section*{Usage Notes}


\begin{figure}[t]
    \centering
    \includegraphics[width=\textwidth]{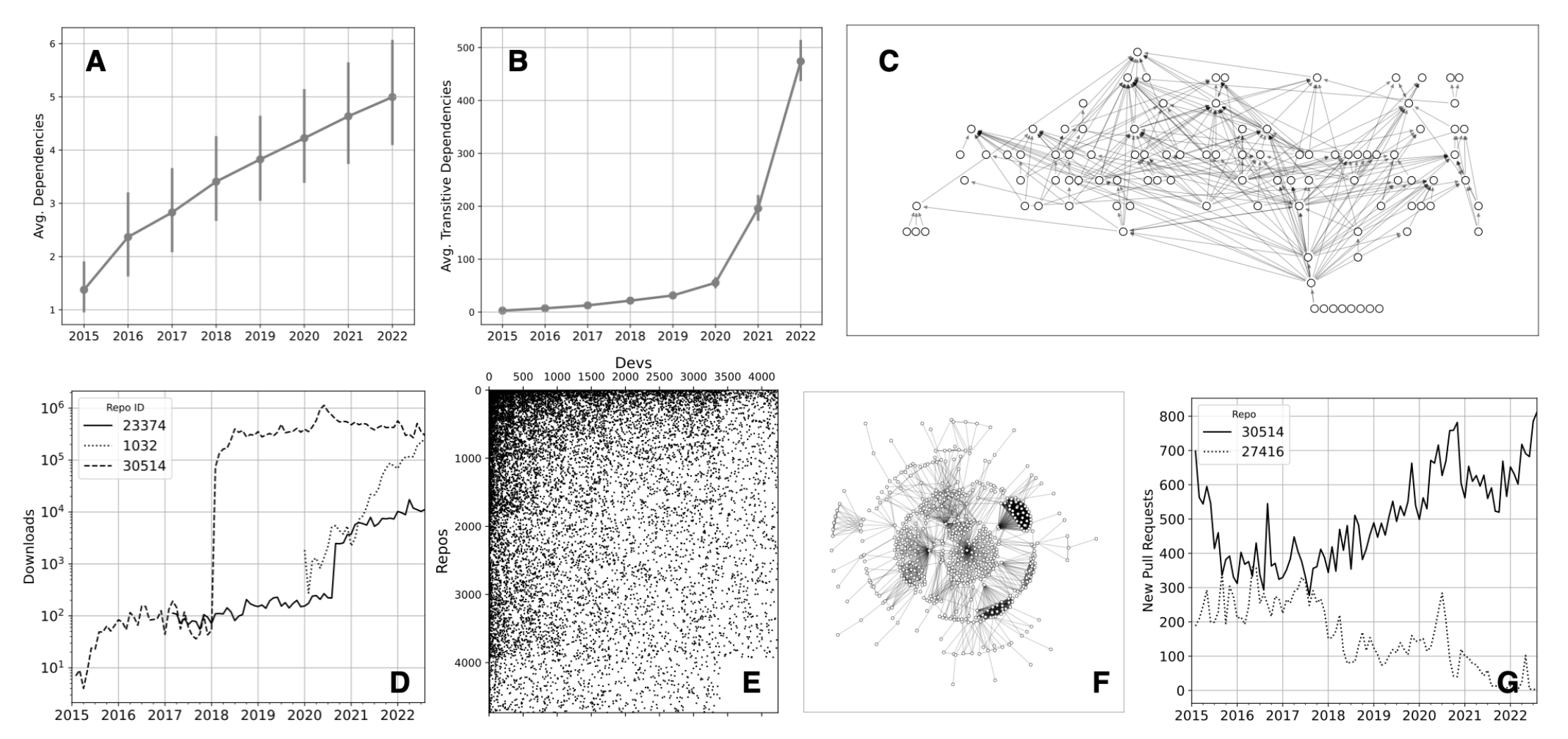}
    \caption{Illustrative plots from demonstration notebooks indicating potential data-processing workflows. A) Average number of dependencies per package at the beginning of each year. B) Evolution of the number of transitive dependencies per package. C) The dependency network of the 100 most downloaded Rust packages in September 2022. D) Time series of monthly downloads of three successful Rust packages. E) Bipartite adjacency matrix of users/developers and the repos they work on in the year 2021, lightly filtered. F) Developer-developer collaboration network in 2021, filtered for developers collaborating on at least three repos. G) Monthly time series of new pull requests to two initially successful Rust libraries.}
    \label{fig:demos}
\end{figure}

To demonstrate how to read in and analyse the database, we provide short Jupyter notebooks that extract data and carry out elementary data manipulations, with the data aggregated at the monthly level. These notebooks are included with the other software in our materials. In one, we create the dependency network of the Rust ecosystem at different times and measure its growth. This data can be used to study ecosystem health, as errors and issues are known to spread through these networks \cite{decan2019empirical}. In another, we plot the time series of downloads and new pull requests to specific repositories over time. Such time series can be used to study the dynamics of collaboration, use, and success at a fine-grained level \cite{sinatra2016quantifying}. In a third, we show how to load in the data of developers and packages as a rectangular matrix, which can be analysed as a bipartite network or, after a projection, as a developer-developer collaboration network. The bipartite network could be used to study the overall complexity of the ecosystem \cite{hidalgo2009building,servedio2018new}, while the developer network reveals patterns of collaborations between projects \cite{singh2010small}. In Figure~\ref{fig:demos} we present several illustrative examples of descriptive analyses resulting from the demonstration notebooks. 

Our dataset can also be used as an input to study current themes of the empirical software engineering community. For example, our data can extend in time the work of Decan et al. \cite{decan2019empirical} on transitive dependencies and library centrality in Rust. These measures themselves provide interesting ways to quantify success and importance of libraries.

Another interesting area of research that our data might be applied to is the concept of code smells, in particular community smells \cite{tamburri2019exploring}. Code smells are problematic kinds of patterns in software, while community smells refer to such sub-optimal patterns on the social and collaborative levels of collective software development. For example, Tamburri et al. \cite{tamburri2019exploring} describe several community smells such as the ``lone wolf'' - when a single developer acts in a unilateral and inconsiderate way - or the ``organizational silo'' - when developer teams working on different parts of a codebase only communicate through one or two team members. These smells are quantified in part by considering the collaboration and communication networks of developers. Our dataset can be used to calculate many of these concepts including collaboration network measures and socio-technical congruence \cite{cataldo2008socio} at the ecosystem level by considering which developers contribute to which libraries and their interactions in dealing with issues and pull requests. Social outcomes like turnover, which are used to test and validate measures of community smells, can also easily be measured.

A third line of software engineering research which our data can complement is the question of use and success of software. As mentioned earlier, GitHub stars are imperfect indicators of successful or high-quality software \cite{borges2018s}. Recent work has sought to refine measure of software success by studying why users adopt specific software \cite{li2022exploring}. In other words: what factors and metrics do users take into account when picking an OSS solution for a problem? Beyond the metrics reported in our data such as stars, downloads, and watchers, many others mentioned in this literature can be calculated. These include metrics of community support and adoption (number of contributors, issue and pull request response times) and maturity (releases, number of forks, age), and to some extent quality (code size and rate of issue resolution).

More generally, that is beyond the broad research field of empirical software engineering, our dataset can be used to explore the interactions between social collaboration, technical dependencies, and the visibility and usage of components of a large software system. The dynamic interactions between these layers of the data offer significant potential for research relating to massive decentralized cooperation (similar to Wikipedia \cite{brandes2009network,mestyan2013early}), the dynamics of teams and their success in digital communities \cite{klug2016understanding}, and the evolution of software systems \cite{sole2020evolving}.

\section*{Code availability}
Code to recreate the database is included in our Figshare \cite{ws2022data} upload: \url{hhttps://doi.org/10.6084/m9.figshare.c.5983534.v1}, and can also be found in a dedicated repository \url{https://github.com/wschuell/repo_datasets}. The software is written in the Python programming language. The database can be created as either PostgreSQL or SQLite database. Version requirements are recorded in the project's Readme file.

\section*{Acknowledgements}
The extraction and construction of the database has been partially funded by Sony Computer Science Laboratories Paris under a dedicated contract with the Complexity Science Hub Vienna. This work has also been partially supported by the Austrian Research Promotion Agency FFG project \# 882184: CSH-Fortsetzung. The authors thank Am\'elie Desvars-Larrive for advice regarding the manuscript.

\section*{Author contributions statement}
WS, JW, VDPS, VL, ST coordinated the production of the dataset. WS designed the database schema, created the tables, and carried out the dataset validation checks. JW implemented the exploratory analyses and demonstrations. VL and ST supervised and mentored the team.  All authors contributed to writing the data descriptor.

\section*{Competing interests}
The authors declare no competing interests.

\bibliography{bibliography}

\begin{thebibliography}{10}
\urlstyle{rm}
\expandafter\ifx\csname url\endcsname\relax
  \def\url#1{\texttt{#1}}\fi
\expandafter\ifx\csname urlprefix\endcsname\relax\def\urlprefix{URL }\fi
\expandafter\ifx\csname doiprefix\endcsname\relax\def\doiprefix{DOI: }\fi
\providecommand{\bibinfo}[2]{#2}
\providecommand{\eprint}[2][]{\url{#2}}

\bibitem{eghbal2020working}
\bibinfo{author}{Eghbal, N.}
\newblock \emph{\bibinfo{title}{{Working in public: The making and maintenance
  of Open Source Software}}} (\bibinfo{publisher}{Stripe Press},
  \bibinfo{year}{2020}).

\bibitem{lerner2002some}
\bibinfo{author}{Lerner, J.} \& \bibinfo{author}{Tirole, J.}
\newblock \bibinfo{journal}{\bibinfo{title}{Some simple economics of open
  source}}.
\newblock {\emph{\JournalTitle{The Journal of Industrial Economics}}}
  \textbf{\bibinfo{volume}{50}}, \bibinfo{pages}{197--234}
  (\bibinfo{year}{2002}).

\bibitem{greenstein2014digital}
\bibinfo{author}{Greenstein, S.} \& \bibinfo{author}{Nagle, F.}
\newblock \bibinfo{journal}{\bibinfo{title}{{Digital dark matter and the
  economic contribution of Apache}}}.
\newblock {\emph{\JournalTitle{Research Policy}}}
  \textbf{\bibinfo{volume}{43}}, \bibinfo{pages}{623--631}
  (\bibinfo{year}{2014}).

\bibitem{nagle2018learning}
\bibinfo{author}{Nagle, F.}
\newblock \bibinfo{journal}{\bibinfo{title}{Learning by contributing: Gaining
  competitive advantage through contribution to crowdsourced public goods}}.
\newblock {\emph{\JournalTitle{Organization Science}}}
  \textbf{\bibinfo{volume}{29}}, \bibinfo{pages}{569--587}
  (\bibinfo{year}{2018}).

\bibitem{nagle2019open}
\bibinfo{author}{Nagle, F.}
\newblock \bibinfo{journal}{\bibinfo{title}{{Open Source Software and firm
  productivity}}}.
\newblock {\emph{\JournalTitle{Management Science}}}
  \textbf{\bibinfo{volume}{65}}, \bibinfo{pages}{1191--1215}
  (\bibinfo{year}{2019}).

\bibitem{decan2019empirical}
\bibinfo{author}{Decan, A.}, \bibinfo{author}{Mens, T.} \&
  \bibinfo{author}{Grosjean, P.}
\newblock \bibinfo{journal}{\bibinfo{title}{An empirical comparison of
  dependency network evolution in seven software packaging ecosystems}}.
\newblock {\emph{\JournalTitle{Empirical Software Engineering}}}
  \textbf{\bibinfo{volume}{24}}, \bibinfo{pages}{381--416}
  (\bibinfo{year}{2019}).

\bibitem{zoller2020topology}
\bibinfo{author}{Z{\"o}ller, N.}, \bibinfo{author}{Morgan, J.~H.} \&
  \bibinfo{author}{Schr{\"o}der, T.}
\newblock \bibinfo{journal}{\bibinfo{title}{A topology of groups: What github
  can tell us about online collaboration}}.
\newblock {\emph{\JournalTitle{Technological Forecasting and Social Change}}}
  \textbf{\bibinfo{volume}{161}}, \bibinfo{pages}{120291}
  (\bibinfo{year}{2020}).

\bibitem{sinatra2016quantifying}
\bibinfo{author}{Sinatra, R.}, \bibinfo{author}{Wang, D.},
  \bibinfo{author}{Deville, P.}, \bibinfo{author}{Song, C.} \&
  \bibinfo{author}{Barab{\'a}si, A.-L.}
\newblock \bibinfo{journal}{\bibinfo{title}{Quantifying the evolution of
  individual scientific impact}}.
\newblock {\emph{\JournalTitle{Science}}} \textbf{\bibinfo{volume}{354}},
  \bibinfo{pages}{aaf5239} (\bibinfo{year}{2016}).

\bibitem{cataldo2008socio}
\bibinfo{author}{Cataldo, M.}, \bibinfo{author}{Herbsleb, J.~D.} \&
  \bibinfo{author}{Carley, K.~M.}
\newblock \bibinfo{title}{Socio-technical congruence: a framework for assessing
  the impact of technical and work dependencies on software development
  productivity}.
\newblock In \emph{\bibinfo{booktitle}{{Proceedings of the Second ACM-IEEE
  international symposium on Empirical Software Engineering and Measurement
  (ESEM)}}}, \bibinfo{pages}{2--11} (\bibinfo{year}{2008}).

\bibitem{jung2021safe}
\bibinfo{author}{Jung, R.}, \bibinfo{author}{Jourdan, J.-H.},
  \bibinfo{author}{Krebbers, R.} \& \bibinfo{author}{Dreyer, D.}
\newblock \bibinfo{journal}{\bibinfo{title}{Safe systems programming in rust}}.
\newblock {\emph{\JournalTitle{Communications of the ACM}}}
  \textbf{\bibinfo{volume}{64}}, \bibinfo{pages}{144--152}
  (\bibinfo{year}{2021}).

\bibitem{kalliamvakou2016depth}
\bibinfo{author}{Kalliamvakou, E.} \emph{et~al.}
\newblock \bibinfo{journal}{\bibinfo{title}{{An in-depth study of the promises
  and perils of mining GitHub}}}.
\newblock {\emph{\JournalTitle{Empirical Software Engineering}}}
  \textbf{\bibinfo{volume}{21}}, \bibinfo{pages}{2035--2071}
  (\bibinfo{year}{2016}).

\bibitem{fry2020dataset}
\bibinfo{author}{Fry, T.}, \bibinfo{author}{Dey, T.},
  \bibinfo{author}{Karnauch, A.} \& \bibinfo{author}{Mockus, A.}
\newblock \bibinfo{title}{A dataset and an approach for identity resolution of
  38 million author ids extracted from 2b git commits}.
\newblock In \emph{\bibinfo{booktitle}{Proceedings of the 17th international
  conference on mining software repositories}}, \bibinfo{pages}{518--522}
  (\bibinfo{year}{2020}).

\bibitem{gote2021gambit}
\bibinfo{author}{Gote, C.} \& \bibinfo{author}{Zingg, C.}
\newblock \bibinfo{title}{{gambit--An Open Source Name Disambiguation Tool for
  Version Control Systems}}.
\newblock In \emph{\bibinfo{booktitle}{IEEE/ACM 18th International Conference
  on Mining Software Repositories (MSR)}}, \bibinfo{pages}{80--84}
  (\bibinfo{organization}{IEEE}, \bibinfo{year}{2021}).

\bibitem{golzadeh2021ground}
\bibinfo{author}{Golzadeh, M.}, \bibinfo{author}{Decan, A.},
  \bibinfo{author}{Legay, D.} \& \bibinfo{author}{Mens, T.}
\newblock \bibinfo{journal}{\bibinfo{title}{{A ground-truth dataset and
  classification model for detecting bots in GitHub issue and PR comments}}}.
\newblock {\emph{\JournalTitle{Journal of Systems and Software}}}
  \textbf{\bibinfo{volume}{175}}, \bibinfo{pages}{110911}
  (\bibinfo{year}{2021}).

\bibitem{gousios2012ghtorrent}
\bibinfo{author}{Gousios, G.} \& \bibinfo{author}{Spinellis, D.}
\newblock \bibinfo{title}{Ghtorrent: Github's data from a firehose}.
\newblock In \emph{\bibinfo{booktitle}{2012 9th IEEE Working Conference on
  Mining Software Repositories (MSR)}}, \bibinfo{pages}{12--21}
  (\bibinfo{organization}{IEEE}, \bibinfo{year}{2012}).

\bibitem{pietri2019software}
\bibinfo{author}{Pietri, A.}, \bibinfo{author}{Spinellis, D.} \&
  \bibinfo{author}{Zacchiroli, S.}
\newblock \bibinfo{title}{The software heritage graph dataset: public software
  development under one roof}.
\newblock In \emph{\bibinfo{booktitle}{2019 IEEE/ACM 16th International
  Conference on Mining Software Repositories (MSR)}}, \bibinfo{pages}{138--142}
  (\bibinfo{organization}{IEEE}, \bibinfo{year}{2019}).

\bibitem{ma2019world}
\bibinfo{author}{Ma, Y.}, \bibinfo{author}{Bogart, C.},
  \bibinfo{author}{Amreen, S.}, \bibinfo{author}{Zaretzki, R.} \&
  \bibinfo{author}{Mockus, A.}
\newblock \bibinfo{title}{{World of Code: an infrastructure for mining the
  universe of open source VCS data}}.
\newblock In \emph{\bibinfo{booktitle}{2019 IEEE/ACM 16th International
  Conference on Mining Software Repositories (MSR)}}, \bibinfo{pages}{143--154}
  (\bibinfo{organization}{IEEE}, \bibinfo{year}{2019}).

\bibitem{roberts2006understanding}
\bibinfo{author}{Roberts, J.~A.}, \bibinfo{author}{Hann, I.-H.} \&
  \bibinfo{author}{Slaughter, S.~A.}
\newblock \bibinfo{journal}{\bibinfo{title}{{Understanding the motivations,
  participation, and performance of Open Source Software developers: A
  longitudinal study of the Apache projects}}}.
\newblock {\emph{\JournalTitle{Management science}}}
  \textbf{\bibinfo{volume}{52}}, \bibinfo{pages}{984--999}
  (\bibinfo{year}{2006}).

\bibitem{szell2010measuring}
\bibinfo{author}{Szell, M.} \& \bibinfo{author}{Thurner, S.}
\newblock \bibinfo{journal}{\bibinfo{title}{Measuring social dynamics in a
  massive multiplayer online game}}.
\newblock {\emph{\JournalTitle{Social Networks}}}
  \textbf{\bibinfo{volume}{32}}, \bibinfo{pages}{313--329}
  (\bibinfo{year}{2010}).

\bibitem{monechi2019efficient}
\bibinfo{author}{Monechi, B.}, \bibinfo{author}{Pullano, G.} \&
  \bibinfo{author}{Loreto, V.}
\newblock \bibinfo{journal}{\bibinfo{title}{Efficient team structures in an
  open-ended cooperative creativity experiment}}.
\newblock {\emph{\JournalTitle{{Proceedings of the National Academy of
  Sciences}}}} \textbf{\bibinfo{volume}{116}} (\bibinfo{year}{2019}).

\bibitem{klug2016understanding}
\bibinfo{author}{Klug, M.} \& \bibinfo{author}{Bagrow, J.~P.}
\newblock \bibinfo{journal}{\bibinfo{title}{Understanding the group dynamics
  and success of teams}}.
\newblock {\emph{\JournalTitle{{Royal Society Open Science}}}}
  \textbf{\bibinfo{volume}{3}}, \bibinfo{pages}{160007} (\bibinfo{year}{2016}).

\bibitem{ma2018constructing}
\bibinfo{author}{Ma, Y.}
\newblock \bibinfo{title}{{Constructing supply chains in Open Source
  Software}}.
\newblock In \emph{\bibinfo{booktitle}{2018 IEEE/ACM 40th International
  Conference on Software Engineering: Companion (ICSE-Companion)}},
  \bibinfo{pages}{458--459} (\bibinfo{organization}{IEEE},
  \bibinfo{year}{2018}).

\bibitem{zimmermann2019small}
\bibinfo{author}{Zimmermann, M.}, \bibinfo{author}{Staicu, C.-A.},
  \bibinfo{author}{Tenny, C.} \& \bibinfo{author}{Pradel, M.}
\newblock \bibinfo{title}{Small world with high risks: A study of security
  threats in the npm ecosystem}.
\newblock In \emph{\bibinfo{booktitle}{28th USENIX Security Symposium (USENIX
  Security 19)}}, \bibinfo{pages}{995--1010} (\bibinfo{year}{2019}).

\bibitem{ohm2020backstabber}
\bibinfo{author}{Ohm, M.}, \bibinfo{author}{Plate, H.},
  \bibinfo{author}{Sykosch, A.} \& \bibinfo{author}{Meier, M.}
\newblock \bibinfo{title}{{Backstabber’s knife collection: A review of Open
  Source Software supply chain attacks}}.
\newblock In \emph{\bibinfo{booktitle}{International Conference on Detection of
  Intrusions and Malware, and Vulnerability Assessment}},
  \bibinfo{pages}{23--43} (\bibinfo{organization}{Springer},
  \bibinfo{year}{2020}).

\bibitem{decan2019package}
\bibinfo{author}{Decan, A.} \& \bibinfo{author}{Mens, T.}
\newblock \bibinfo{journal}{\bibinfo{title}{What do package dependencies tell
  us about semantic versioning?}}
\newblock {\emph{\JournalTitle{IEEE Transactions on Software Engineering}}}
  \textbf{\bibinfo{volume}{47}}, \bibinfo{pages}{1226--1240}
  (\bibinfo{year}{2019}).

\bibitem{scholtes2016aristotle}
\bibinfo{author}{Scholtes, I.}, \bibinfo{author}{Mavrodiev, P.} \&
  \bibinfo{author}{Schweitzer, F.}
\newblock \bibinfo{journal}{\bibinfo{title}{{From Aristotle to Ringelmann: a
  large-scale analysis of team productivity and coordination in Open Source
  Software projects}}}.
\newblock {\emph{\JournalTitle{Empirical Software Engineering}}}
  \textbf{\bibinfo{volume}{21}}, \bibinfo{pages}{642--683}
  (\bibinfo{year}{2016}).

\bibitem{gote2019git2net}
\bibinfo{author}{Gote, C.}, \bibinfo{author}{Scholtes, I.} \&
  \bibinfo{author}{Schweitzer, F.}
\newblock \bibinfo{title}{git2net-mining time-stamped co-editing networks from
  large git repositories}.
\newblock In \emph{\bibinfo{booktitle}{IEEE/ACM 16th International Conference
  on Mining Software Repositories (MSR)}}, \bibinfo{pages}{433--444}
  (\bibinfo{organization}{IEEE}, \bibinfo{year}{2019}).

\bibitem{dabbish2012social}
\bibinfo{author}{Dabbish, L.}, \bibinfo{author}{Stuart, C.},
  \bibinfo{author}{Tsay, J.} \& \bibinfo{author}{Herbsleb, J.}
\newblock \bibinfo{title}{{Social coding in GitHub: transparency and
  collaboration in an open software repository}}.
\newblock In \emph{\bibinfo{booktitle}{Proceedings of the ACM 2012 conference
  on Computer Supported Cooperative Work}}, \bibinfo{pages}{1277--1286}
  (\bibinfo{year}{2012}).

\bibitem{marlow2013impression}
\bibinfo{author}{Marlow, J.}, \bibinfo{author}{Dabbish, L.} \&
  \bibinfo{author}{Herbsleb, J.}
\newblock \bibinfo{title}{{Impression formation in online peer production:
  activity traces and personal profiles in GitHub}}.
\newblock In \emph{\bibinfo{booktitle}{Proceedings of the 2013 Conference on
  Computer Supported Cooperative Work}}, \bibinfo{pages}{117--128}
  (\bibinfo{year}{2013}).

\bibitem{borges2018s}
\bibinfo{author}{Borges, H.} \& \bibinfo{author}{Valente, M.~T.}
\newblock \bibinfo{journal}{\bibinfo{title}{{What’s in a GitHub star?
  Understanding repository starring practices in a social coding platform}}}.
\newblock {\emph{\JournalTitle{Journal of Systems and Software}}}
  \textbf{\bibinfo{volume}{146}}, \bibinfo{pages}{112--129}
  (\bibinfo{year}{2018}).

\bibitem{moldon2021gamification}
\bibinfo{author}{Moldon, L.}, \bibinfo{author}{Strohmaier, M.} \&
  \bibinfo{author}{Wachs, J.}
\newblock \bibinfo{title}{{How gamification affects software developers:
  Cautionary evidence from a natural experiment on GitHub}}.
\newblock In \emph{\bibinfo{booktitle}{IEEE/ACM 43rd International Conference
  on Software Engineering (ICSE)}}, \bibinfo{pages}{549--561}
  (\bibinfo{organization}{IEEE}, \bibinfo{year}{2021}).

\bibitem{papoutsoglou2019extracting}
\bibinfo{author}{Papoutsoglou, M.}, \bibinfo{author}{Ampatzoglou, A.},
  \bibinfo{author}{Mittas, N.} \& \bibinfo{author}{Angelis, L.}
\newblock \bibinfo{journal}{\bibinfo{title}{Extracting knowledge from on-line
  sources for software engineering labor market: A mapping study}}.
\newblock {\emph{\JournalTitle{IEEE Access}}} \textbf{\bibinfo{volume}{7}},
  \bibinfo{pages}{157595--157613} (\bibinfo{year}{2019}).

\bibitem{shimada2022github}
\bibinfo{author}{Shimada, N.}, \bibinfo{author}{Xiao, T.},
  \bibinfo{author}{Hata, H.}, \bibinfo{author}{Treude, C.} \&
  \bibinfo{author}{Matsumoto, K.}
\newblock \bibinfo{title}{{HGitHub Sponsors: Exploring a New Way to Contribute
  to Open Source}}.
\newblock In \emph{\bibinfo{booktitle}{IEEE/ACM 44th International Conference
  on Software Engineering (ICSE)}} (\bibinfo{organization}{IEEE},
  \bibinfo{year}{2022}).

\bibitem{corominas2013origins}
\bibinfo{author}{Corominas-Murtra, B.}, \bibinfo{author}{Go{\~n}i, J.},
  \bibinfo{author}{Sol{\'e}, R.~V.} \& \bibinfo{author}{Rodr{\'\i}guez-Caso,
  C.}
\newblock \bibinfo{journal}{\bibinfo{title}{On the origins of hierarchy in
  complex networks}}.
\newblock {\emph{\JournalTitle{Proceedings of the National Academy of
  Sciences}}} \textbf{\bibinfo{volume}{110}}, \bibinfo{pages}{13316--13321}
  (\bibinfo{year}{2013}).

\bibitem{montandon2019identifying}
\bibinfo{author}{Montandon, J.~E.}, \bibinfo{author}{Silva, L.~L.} \&
  \bibinfo{author}{Valente, M.~T.}
\newblock \bibinfo{title}{{Identifying experts in software libraries and
  frameworks among GitHub users}}.
\newblock In \emph{\bibinfo{booktitle}{IEEE/ACM 16th International Conference
  on Mining Software Repositories (MSR)}}, \bibinfo{pages}{276--287}
  (\bibinfo{organization}{IEEE}, \bibinfo{year}{2019}).

\bibitem{wessel2018power}
\bibinfo{author}{Wessel, M.} \emph{et~al.}
\newblock \bibinfo{journal}{\bibinfo{title}{{The power of bots: Characterizing
  and understanding bots in OSS projects}}}.
\newblock {\emph{\JournalTitle{Proceedings of the ACM on Human-Computer
  Interaction}}} \textbf{\bibinfo{volume}{2}}, \bibinfo{pages}{1--19}
  (\bibinfo{year}{2018}).

\bibitem{wessel2022bots}
\bibinfo{author}{Wessel, M.} \emph{et~al.}
\newblock \bibinfo{title}{Bots for pull requests: The good, the bad, and the
  promising}.
\newblock In \emph{\bibinfo{booktitle}{44th ACM/IEEE International Conference
  on Software Engineering (ICSE)}}, vol.~\bibinfo{volume}{26},
  \bibinfo{pages}{16} (\bibinfo{organization}{ACM/IEEE}, \bibinfo{year}{2022}).

\bibitem{ws2022data}
\bibinfo{author}{Schueller, W.}, \bibinfo{author}{Wachs, J.},
  \bibinfo{author}{Servedio, V.~D.}, \bibinfo{author}{Thurner, S.} \&
  \bibinfo{author}{Loreto, V.}
\newblock \bibinfo{journal}{\bibinfo{title}{{Replication Data for Evolving
  collaboration, dependencies, and use in the Rust Open Source Software
  ecosystem}}}.
\newblock {\emph{\JournalTitle{figshare}}}
  \url{https://doi.org/10.6084/m9.figshare.c.5983534.v1}
  (\bibinfo{year}{2022}).

\bibitem{vasilescu2015gender}
\bibinfo{author}{Vasilescu, B.} \emph{et~al.}
\newblock \bibinfo{title}{Gender and tenure diversity in github teams}.
\newblock In \emph{\bibinfo{booktitle}{Proceedings of the 33rd annual ACM
  conference on human factors in computing systems}},
  \bibinfo{pages}{3789--3798} (\bibinfo{year}{2015}).

\bibitem{rossi2022gender}
\bibinfo{author}{Rossi, D.} \& \bibinfo{author}{Zacchiroli, S.}
\newblock \bibinfo{title}{Worldwide gender differences in public code
  contributions: and how they have been affected by the covid-19 pandemic}.
\newblock In \emph{\bibinfo{booktitle}{IEEE/ACM 44th International Conference
  on Software Engineering (ICSE)}} (\bibinfo{year}{2022}).

\bibitem{rastogi2018relationship}
\bibinfo{author}{Rastogi, A.}, \bibinfo{author}{Nagappan, N.},
  \bibinfo{author}{Gousios, G.} \& \bibinfo{author}{van~der Hoek, A.}
\newblock \bibinfo{title}{{Relationship between geographical location and
  evaluation of developer contributions in GitHub}}.
\newblock In \emph{\bibinfo{booktitle}{{Proceedings of the 12th ACM/IEEE
  International Symposium on Empirical Software Engineering and Measurement
  (ESEM)}}}, \bibinfo{pages}{1--8} (\bibinfo{year}{2018}).

\bibitem{braesemann2019global}
\bibinfo{author}{Braesemann, F.}, \bibinfo{author}{Stoehr, N.} \&
  \bibinfo{author}{Graham, M.}
\newblock \bibinfo{journal}{\bibinfo{title}{Global networks in collaborative
  programming}}.
\newblock {\emph{\JournalTitle{Regional Studies, Regional Science}}}
  \textbf{\bibinfo{volume}{6}}, \bibinfo{pages}{371--373}
  (\bibinfo{year}{2019}).

\bibitem{wachs2022geography}
\bibinfo{author}{Wachs, J.}, \bibinfo{author}{Nitecki, M.},
  \bibinfo{author}{Schueller, W.} \& \bibinfo{author}{Polleres, A.}
\newblock \bibinfo{journal}{\bibinfo{title}{{The Geography of Open Source
  Software: Evidence from GitHub}}}.
\newblock {\emph{\JournalTitle{Technological Forecasting and Social Change}}}
  \textbf{\bibinfo{volume}{176}}, \bibinfo{pages}{121478}
  (\bibinfo{year}{2022}).

\bibitem{prana2021including}
\bibinfo{author}{Prana, G. A.~A.} \emph{et~al.}
\newblock \bibinfo{journal}{\bibinfo{title}{{Including everyone, everywhere:
  Understanding opportunities and challenges of geographic gender-inclusion in
  OSS}}}.
\newblock {\emph{\JournalTitle{IEEE Transactions on Software Engineering}}}
  (\bibinfo{year}{2021}).

\bibitem{gousios2017mining}
\bibinfo{author}{Gousios, G.} \& \bibinfo{author}{Spinellis, D.}
\newblock \bibinfo{title}{{Mining software engineering data from GitHub}}.
\newblock In \emph{\bibinfo{booktitle}{IEEE/ACM 39th International Conference
  on Software Engineering Companion (ICSE-C)}}, \bibinfo{pages}{501--502}
  (\bibinfo{organization}{IEEE}, \bibinfo{year}{2017}).

\bibitem{gonzalez2022software}
\bibinfo{author}{Gonzalez-Barahona, J.~M.},
  \bibinfo{author}{Izquierdo-Cort{\'a}zar, D.} \& \bibinfo{author}{Robles, G.}
\newblock \bibinfo{journal}{\bibinfo{title}{Software development metrics with a
  purpose}}.
\newblock {\emph{\JournalTitle{Computer}}} \textbf{\bibinfo{volume}{55}},
  \bibinfo{pages}{66--73} (\bibinfo{year}{2022}).

\bibitem{hidalgo2009building}
\bibinfo{author}{Hidalgo, C.~A.} \& \bibinfo{author}{Hausmann, R.}
\newblock \bibinfo{journal}{\bibinfo{title}{The building blocks of economic
  complexity}}.
\newblock {\emph{\JournalTitle{Proceedings of the national academy of
  sciences}}} \textbf{\bibinfo{volume}{106}}, \bibinfo{pages}{10570--10575}
  (\bibinfo{year}{2009}).

\bibitem{servedio2018new}
\bibinfo{author}{Servedio, V. D.~P.}, \bibinfo{author}{Butt{\`a}, P.},
  \bibinfo{author}{Mazzilli, D.}, \bibinfo{author}{Tacchella, A.} \&
  \bibinfo{author}{Pietronero, L.}
\newblock \bibinfo{journal}{\bibinfo{title}{A new and stable estimation method
  of country economic fitness and product complexity}}.
\newblock {\emph{\JournalTitle{Entropy}}} \textbf{\bibinfo{volume}{20}},
  \bibinfo{pages}{783} (\bibinfo{year}{2018}).

\bibitem{singh2010small}
\bibinfo{author}{Singh, P.~V.}
\newblock \bibinfo{journal}{\bibinfo{title}{The small-world effect: The
  influence of macro-level properties of developer collaboration networks on
  open-source project success}}.
\newblock {\emph{\JournalTitle{ACM Transactions on Software Engineering and
  Methodology (TOSEM)}}} \textbf{\bibinfo{volume}{20}}, \bibinfo{pages}{1--27}
  (\bibinfo{year}{2010}).

\bibitem{tamburri2019exploring}
\bibinfo{author}{Tamburri, D.~A.}, \bibinfo{author}{Palomba, F.} \&
  \bibinfo{author}{Kazman, R.}
\newblock \bibinfo{journal}{\bibinfo{title}{Exploring community smells in
  open-source: An automated approach}}.
\newblock {\emph{\JournalTitle{IEEE Transactions on software Engineering}}}
  \textbf{\bibinfo{volume}{47}}, \bibinfo{pages}{630--652}
  (\bibinfo{year}{2019}).

\bibitem{li2022exploring}
\bibinfo{author}{Li, X.}, \bibinfo{author}{Moreschini, S.},
  \bibinfo{author}{Zhang, Z.} \& \bibinfo{author}{Taibi, D.}
\newblock \bibinfo{journal}{\bibinfo{title}{{Exploring factors and metrics to
  select Open Source Software components for integration: An empirical
  study}}}.
\newblock {\emph{\JournalTitle{Journal of Systems and Software}}}
  \textbf{\bibinfo{volume}{188}}, \bibinfo{pages}{111255}
  (\bibinfo{year}{2022}).

\bibitem{brandes2009network}
\bibinfo{author}{Brandes, U.}, \bibinfo{author}{Kenis, P.},
  \bibinfo{author}{Lerner, J.} \& \bibinfo{author}{Van~Raaij, D.}
\newblock \bibinfo{title}{Network analysis of collaboration structure in
  wikipedia}.
\newblock In \emph{\bibinfo{booktitle}{Proceedings of the 18th international
  conference on World wide web}}, \bibinfo{pages}{731--740}
  (\bibinfo{year}{2009}).

\bibitem{mestyan2013early}
\bibinfo{author}{Mesty{\'a}n, M.}, \bibinfo{author}{Yasseri, T.} \&
  \bibinfo{author}{Kert{\'e}sz, J.}
\newblock \bibinfo{journal}{\bibinfo{title}{Early prediction of movie box
  office success based on wikipedia activity big data}}.
\newblock {\emph{\JournalTitle{PloS one}}} \textbf{\bibinfo{volume}{8}},
  \bibinfo{pages}{e71226} (\bibinfo{year}{2013}).

\bibitem{sole2020evolving}
\bibinfo{author}{Sol{\'e}, R.} \& \bibinfo{author}{Valverde, S.}
\newblock \bibinfo{journal}{\bibinfo{title}{Evolving complexity: how tinkering
  shapes cells, software and ecological networks}}.
\newblock {\emph{\JournalTitle{Philosophical Transactions of the Royal Society
  B}}} \textbf{\bibinfo{volume}{375}}, \bibinfo{pages}{20190325}
  (\bibinfo{year}{2020}).

\end{thebibliography}

\end{document}